# OMICtools: a community-driven search engine for biological data analysis

Helene Perrin[1,2], Marion Denorme[1], Julien Grosjean[3], OMICtools community[4], Emeric Dynomant[1], Vincent J. Henry[2], Fabien Pichon[1], Stefan Darmoni[3], Arnaud Desfeux[1], Bruno J. Gonzalez[2]*

[1]omicX, Seine Innopolis, 72 rue de la republique, 76140 Le-Petit-Quevilly, France, [2]Normandie Univ, UNIROUEN, Inserm U1245 and Rouen University Hospital, Normandy Center for Genomic and Personalized Medicine, Rouen, France, [3]CISMeF, TIBS, LITIS EA 4108, Rouen University Hospital, rue de Germont, Rouen, France; UMR_S 1142, LIMICS, Inserm, Paris, France, [4] https://omictools.com/community.
*Corresponding author: Tel: +33 2 35 14 85 47. Email: bruno.gonzales@univ-rouen.fr

**ABSTRACT**

With high-throughput biotechnologies generating unprecedented quantities of data, researchers are faced with the challenge of locating and comparing an exponentially growing number of programs and websites dedicated to computational biology, in order to maximize the potential of their data. OMICtools is designed to meet this need with its open-access search engine offering an easy means of locating the right tools corresponding to each researcher and their specific biological data analyses. The OMICtools website (https://OMICtools.com) centralizes more than 18,500 software tools and databases, manually classified, by a team of biocurators including many scientific experts. Key information, a direct link, and access to discussions and evaluations by the biomedical community are provided for each tool. Anyone can join the OMICtools community and create a profile page, to share their expertise and comment on tools. In addition, developers can directly upload their code versions which are registered for identification and citation in scientific publications, improving research traceability. The OMICtools community links thousands of life scientists and developers worldwide, who use this bioinformatics platform to accelerate their projects and biological data analyses.

# Introduction

The last two decades have seen rapid advances in high-throughput biotechnologies, such as next-generation sequencing (NGS), microarray, mass spectrometry and nuclear magnetic resonance, all of which produce vast amounts of complex and multi-disciplinary data (1,2,3). Computational science expertise is essential to manage these big data, to this end a myriad of databases, methods and software tools have been developed (4,5,6). Among them, biologists and clinicians have to choose an appropriate tool set that matches both their scientific aims and their computer skills, in order to optimally analyze the different steps of their experiments and to extract biologically relevant information (7,8). Yet bioinformatic tools and databases are sadly too often widely scattered throughout the massive quantity of publications and information available on the web, and locating them presents ongoing challenges. There is now an urgent need to organize the bioinformatic resources and to make them more accessible for researchers.

Only a few projects, ensured by the combined efforts of some institutes, have attempted to meet this challenge. Webbased free platforms have been created to help researchers perform, store and share their data and analyses, such as the Galaxy project (https://galaxyproject.org) dedicated to genomics (9) and the iPlant Collaborative cyberinfrastructure (www.cyverse.org) for plant, animal and microbial genome datasets (10). Created in 2015, the European infrastructure for biological information ELIXIR (https://bio.tools) offers a list of about 1,800 tools, provided by developers and scientists from various institutions (11) who thereby develop the underlying bioinformatics-specific ontology EDAM (12). Knowledge-sharing is also promoted in some open-access forums such as BioStar (http://biostars.org) or SEQanswers wiki (http://SEQanswers.com) where scientists can ask the bioinformatics community technical questions about particular tools dedicated to genomics and NGS (13). Nonetheless, finding the right solution remains challenging for many biologists, especially for those with minimal computing experience. To date, available resources are limited to particular domains of life sciences and offer relatively short lists of specific tools. Furthermore, registry curation and documentation rely fully on the efforts of program developers and researchers, which does not allow tracking of updates or discontinuations of software tools and databases. A novel approach providing a continuously updated overview of both existing and newly available highly diverse bioinformatic resources is urgently needed, along with better support to allow all researchers to use these resources easily.

OMICtools (https://OMICtools.com), an innovative open-access platform offering an intuitive search experience to retrieve biological data analysis tools, was developed to fulfil this need. Since our first report in 2014 (14), we have significantly upgraded the scale and quality of the OMICtools website. The tool set has been increased by more than 350%, to reach over 18,500 available in June 2017. OMICtools biocurators continuously collect information from original articles, websites and bioinformatic repositories to make available on the site the latest analytical software and databases. Pertinent detailed information and a direct link, as well as evaluations by the biomedical research community are provided for each tool.

We are currently adding about 1,000 new resources and updating 250 existing tools each month - and these numbers are growing constantly. Users can either browse freely the didactic catalog or use the new powerful search engine to directly find the tools they need. New website features allow unrestricted exchange of tool feedback and ideas with other users, as well as asking questions to the tool developers directly. Developers can also upload their source codes to the site, thereby making them citable by unique identifiers. They can also collaborate within the OMICtools community when developing new tools and maintain existing tools up-to-date. With a current mean of 80,000 visits per month worldwide, over 4600 registered users and 72 expert biocurators, OMICtools is playing a leading role in the development of a new interactive community of biologists and bioinformaticians, dedicated to biotechnology, data analysis and scientific innovation.

## A search engine for choosing the right biological data analysis tools

The OMICtools website (https://omictools.com) provides a list of more than 18,200 web-accessible software tools and databases, all annotated and methodically organized in a three-level classification covering biotechnologies, interpretation or biological topics (e.g., high-throughput sequencing, proteomics, or drug discovery) through to the precise steps of analysis, such as read quality control for whole-genome sequencing, or secondary structure prediction software tools for protein structure analysis. A visual system encompasses icons designed to symbolize each of the categories and a specific color assigned to indicate the associated omics field (i.e. genomics, epigenomics, transcriptomics, proteomics, phenomics, metabolomics, etc.), simplifying user navigation on the site. We recently greatly improved the usability of the OMICtools website with the addition of a powerful search engine. This interactive query interface offers a user-friendly method for researchers to find the tools they need, regardless of their level of expertise in bioinformatics or high-throughput technologies.

As part of the development of the search engine, we constructed a terminology which encompasses the OMICtools' categories and sub-categories as well as an extensive list of synonyms, keywords, abbreviations and expressions commonly used by the scientific community. Our terminology currently comprises 1,151 concepts and more than 16,000 terms, which are linked semantically to ensure consistency and access to related information. Each tool classified on the website is indexed according to this terminology. The OMICtools search engine is based on a REST web service, developed by the CISMeF team (*Catalogue et Index des Sites Medicaux de langue Française* (http://www.churouen.fr/cismef) which has NoSQL access to the OMICtools MySQL database (**Supplemental figure 1**). Queries can employ natural language, Boolean grammar or specific grammar such as a tool acronym or OMICtools category identifier. Using Lucene Java frameworks and related algorithms, the search engine performs a full-text search, even if more than one concept or category of the OMICtools terminology have been identified. When querying the index, the search engine also takes subsumption hierarchy into consideration, to retrieve the tools indexed within the descendant subcategories. Compared with similar search engines, the OMICtools search engine has optimized speed, precision and recall performances. It rapidly identifies a list of relevant tools matching a query using vocabulary from the specific field or common expression. In addition, filters are proposed to make searches more precise. Users can select the OMICtools category, the operating system (Linux, MacOS, Windows, etc…), the programing languages (Python, Java, C++, etc…), the interface (command Line, web user or graphic user interfaces), and the technology.

Moreover, the search engine attributes weights to several parameters for tool ranking that depend principally on its domains of application, which is reflected in the tools's classification in OMICtools. Tool ranking also takes into account the availability and quality of the website functional link, documentation, support, tutorial, etc.), the number and quality of associated publications, and OMICtools users' rating and comments. Because too often the most used tools tend to be the first tools developed but are not necessarily the best ones available (15), our ranking strategy is designed to emphasize tool quality, allowing new effective and promising tools to be highlighted. Once the user has chosen a tool, we systematically propose related tools within the same category that are potentially of interest to the researcher (**Supplemental figure 2**).

## An interactive community to improve knowledge and skills in bioinformatics

OMICtools motivates the development of a new interactive community of biologists and program developers dedicated to high-throughput technologies, data analysis and innovation in biology and medicine (**Figure 1**). Visitors are strongly encouraged to create a free account to join the OMICtools community. This way, users can connect with each other to share and discuss bioinformatic tools.

OMICtools users can add a biography to their personal profile page presenting their educational background and work experience, add information and a link to their institute or company, as well as connect their OMICtools page with their own twitter and ORCID accounts (https://orcid.org/). A bookmarking system allows users to save their favorite tools, organize them in personal collections or share them with other users. Bookmarks can be directly accessed via the user profile page, and OMICtools sends an email notification when users' favorite tools are updated.

Among the more than 4,500 users who have created an OMICtools account, 72 participate in the OMICtools registry effort as external biocurators in their domain of interest. As a biocurator, a user can submit and edit tools, propose categories, review the tools, offer user support and highlight any tools issues, thereby contributing to and managing their own domain of interest. Information about biocurators is available on the community page and on the page of categories in their field of expertise. OMICtools encourages scientists to become biocurators in their fields of expertise. While only minimal information is required to submit a new tool (tool name, description, website URL and webmaster's email address), submitters are encouraged to detail specifications and link references. Submissions are reviewed by OMICtools' data scientists and a complete tool card for the new tool is available within a few days.

An interface has been developed to allow users to report problems, add comments, ask questions and share their reviews and tool ratings with the community. To further encourage scientists and developers to document their resource and interact with the community, a wiki mode allowing user editing has been implemented (**Supplemental figure 2**). Users log into their account to access the option to edit and modify information of any tool page. By joining the OMICtools community, every user can be an actor in developing their own field of expertise and benefit from new contacts, thereby promoting their progress in biological data analysis, and gaining new skills and ideas for developing projects.

To improve recognition and highlight bioinformaticians' skills, OMICtools has implemented the Contributor Roles Taxonomy system, CRediT (http://docs.casrai.org/CRediT). Built on the work of the Wellcome Trust and Harvard University in 2012, this project brings together biomedical publishers, and academic and funding communities to encourage scientists to clearly identify their contribution to collaborative projects. This digital taxonomy includes 14 badges to indicate contributor-ship, such as study conception, manuscript preparation, data curation, computation, etc… (16). On the OMICtools website, users can assign one or more badge to indicate their contributory roles in the development of a specific bioinformatics tools. Badges are shown under the publication associated with the tool on both the tool page and the user profile. The benefits for developers include not only increased professional exposure, and therefore use of their programs, but also the development of connections with users who can provide feedback on bugs and make suggestions for improvements.

The OMICtools community is linking expert biocurators and program developers who submit and categorize tools and maintain resources up-to-date, to users who strengthen the interface by bringing feedback and reviews.

## A tool repository platform to ensure better quality and reproducibility of computational analysis

For unambiguous identification of bioinformatic tools and to make them citable in scientific publications, OMICtools registers tools by two unique identifiers. Firstly, for each software and database within the website, OMICtools adds a unique Research Resource Identifier (RRID), developed under the Resource Identification Initiative, which is transferred to the Neuroscience Information Framework (NIF) registry (https://www.force11.org/group/resource-identificationinitiative). In addition, each source code version uploaded on the OMICtools website is also registered with a digital object identifier (DOI) (https://www.doi.org/). Indeed, developers can directly upload their software tools into the OMICtools server. They indicate the version of the source code, the operating system and architecture, as well as the publication to link to the code. Programmatic access to DataCite's API automatically generates the corresponding DOI. This unique alphanumeric string is defined by the International DOI Foundation and assigned by OMICtools to precisely identify content and provide a persistent link to its location on the web. Uploaded tools are made available on the OMICtools website where users can easily locate them via the search engine. Using OMICtools, developers can modify and update their own work and also benefit from feedback from the OMICtools community to improve and deploy their codes.

These services are designed to facilitate the development, maintenance and follow-up of bioinformatic tools by the authors themselves. Thereby, OMICtools promotes the citation of bioinformatics resources in the scientific literature to ensure exact tool identification and reproducibility and traceability of biological data analysis.

## Discussion

Biological analyses are becoming more and more complex due to the increasing number of steps required to process the raw data and the exponentially growing bioinformatic tools that can be used in each step. Addressing a broad range of life science technologies and topics, OMICtools aims to support researchers and clinicians in navigating the thousands of software tools and databases available. This is the first open-access search engine fully dedicated to biological data analysis. Since our first report in 2014 (14), several new website features facilitate our visitors in their use of the website and by consequence in their choice of bioinformatics tools. Users can access more than 18,500 software tools and databases, each with a complete manually curated description and regularly updated links.

OMICTools website's organization and design offer ease of use. The search engine, with the help of filters, rapidly returns a list of the most relevant tools corresponding to the user's needs. Moreover, reviews and ranking from other scientists help users to choose from these tools. OMICtools promotes the development of a community for bioinformaticians and biologists from various fields to interact and work together. Biologists can save precious time by using the OMICtools platform to guide them through the various steps of their bioinformatics analyzes. By using OMICtools, developers do not waste time developing existing tools. It offers a practical, applicable and straightforward user-friendly platform for programmers to publish and maintain their tools up-to-date. By using this approach, OMICtools not only favors the recognition and follow-up of bioinformatics studies but also helps to improve biological data analysis (17).

Scientific quality and reproducibility largely depend on the standardization and traceability of data and methods used to generate results. This is particularly critical for "omics" approaches but is difficult to attain when using such diverse highthroughput technologies (18). OMICtools has directly addressed this issue by developing several strategies. First, citations and references are specified for each of the 18,500 tools, so that scientists can easily cite methods and software they used. The quality of the tool links is a critical parameter in tool ranking. An automatic link checker assists biocurators identify obsolete links. The successive versions of the tools and obsolete links are indicated but not deleted to facilitate identification and reproducibility of bioinformatic analyses. Moreover, the RRID assigned to each analytical tool and database in the website in collaboration with the NIF registry helps researchers cite the resources used to produce the scientific findings reported in the biomedical literature (19). Finally, programmers can directly upload their tools to the OMICtools website and each published source code version is registered with a unique DOI which provides a persistent identification of bioinformatic resources, even if material is moved or rearranged and an interoperable exchange with other resources on digital networks. Hence OMICtools, which aims to support scientists in their analyses and understanding of their biological datasets, also improves the precision of citing bioinformatics methods used to produce and reproduce results, thereby promoting the quality of scientific publications.

Taken together, OMICtools is a unique interactive and collaborative platform that centralizes all relevant information for searching, reviewing, developing and follow-up of bioinformatics tools. OMICtools brings together an interactive worldwide community of expert biocurators, program developers and biologists, allowing researchers worldwide to take full advantage of new biotechnologies and big data and contribute to quality, speed and progress of biomedical research.

**FUNDING**

This work was supported by the Normandy valorization, Rouen university, INSERM, Bpifrance, foundation Nelia and et Amadeo Barletta (FNAB) and Oncology – Therapeutic, Development (O.T.D). H.P. received a fellowship from Normandy University.

**ACKNOWLEDGEMENTS**

We thank the OMICtools team: Sarah Mackenzie, Nicolas Jacquet, Angelo Marhouchi, Simon Fihue, Enis Boughanmi, Ulrich Moutoussamy, Christal Chambost, Marine Le Mercier, Christophe Mazars, Cyrille Petit, Emeline Duquenne, Elhadji Thioune, Fanny Hardier, Lea Segas, David Tixier, Aminata Sene, Emilie Gondouin, and all the OMICtools users, for their helpful ongoing collaboration in providing high-quality and updated information on bioinformatics tools.

## FIGURE LEGENDS

**Figure 1: OMICtools community driven platform for bioinformatics**

Overview of the collaborative functionality offers on OMICtools platform.
Any user can join the OMICtools community to share its own expertise and acquire new skills. Each of the software tools and databases are identified by a RRID number, its descriptive page can be modified by registered user through a wiki mode. Users and expert biocurators can submit new tools, add reviews and ranking. They can save their favorite tools in collections that they can share with other users. User's profile page can be linked to his Twitter and ORCID accounts. Tools are associated with publication(s) (identified by doi and PMID numbers from Pubmed) and user can indicate its contributor-ship using the CRediT badges. OMICtools is also associated to linked'In and Research Gate. Through OMICtools GitLab, developers can work together and directly upload their tools which get registered with a doi for identification and citation in scientific publications.

## SUPPLEMENTARY FIGURES

Supplementary Data are available at NAR online.

**Supplemental Figure 1: OMICtools search engine**
The OMICtools search engine is based on a REST web service, developed by the CISMeF team (*Catalogue et Index des Sites Medicaux de langue Française,* (http://www.chu-rouen.fr/cismef) who has a NoSQL access to OMICtools MySQL database. Using Lucene Java frameworks and related algorithms, the search engine performs a full-text search in the query. It returns in seconds a list of relevant tools matching with the OMICtools termino-ontology which include all the OMICtools' categories and sub-categories as well as an extensive list of synonyms, keywords, abbreviations and expressions commonly used by the biologist's community.

**Supplemental Figure 2: Tool page on OMICtools.**
Screenshot from the OMICtools websie. Exemple of SAMtools page (https://omictools.com/samtools-tool) and tutorial.

**Figure 1**

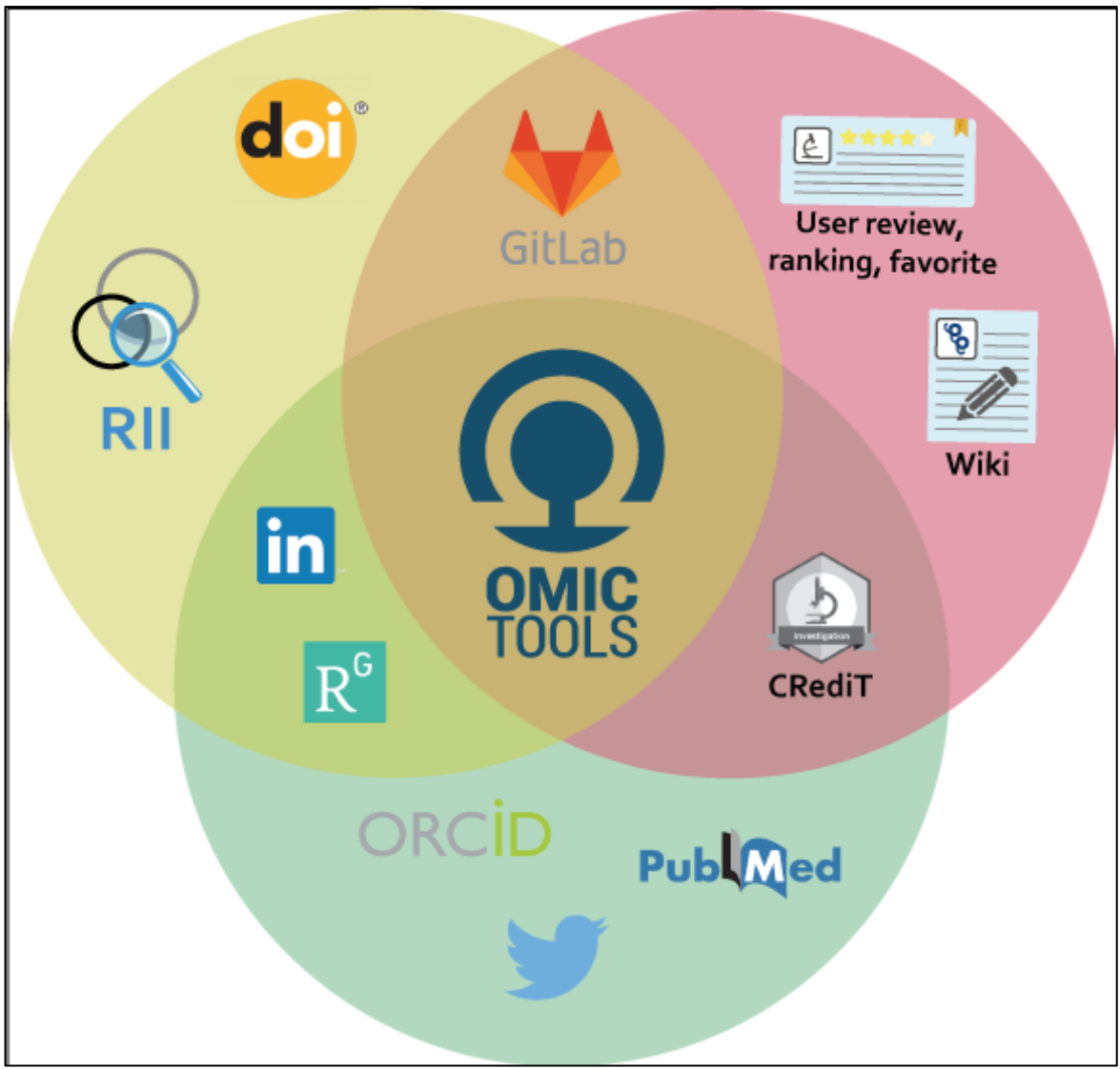

**Suppl. Figure 1**

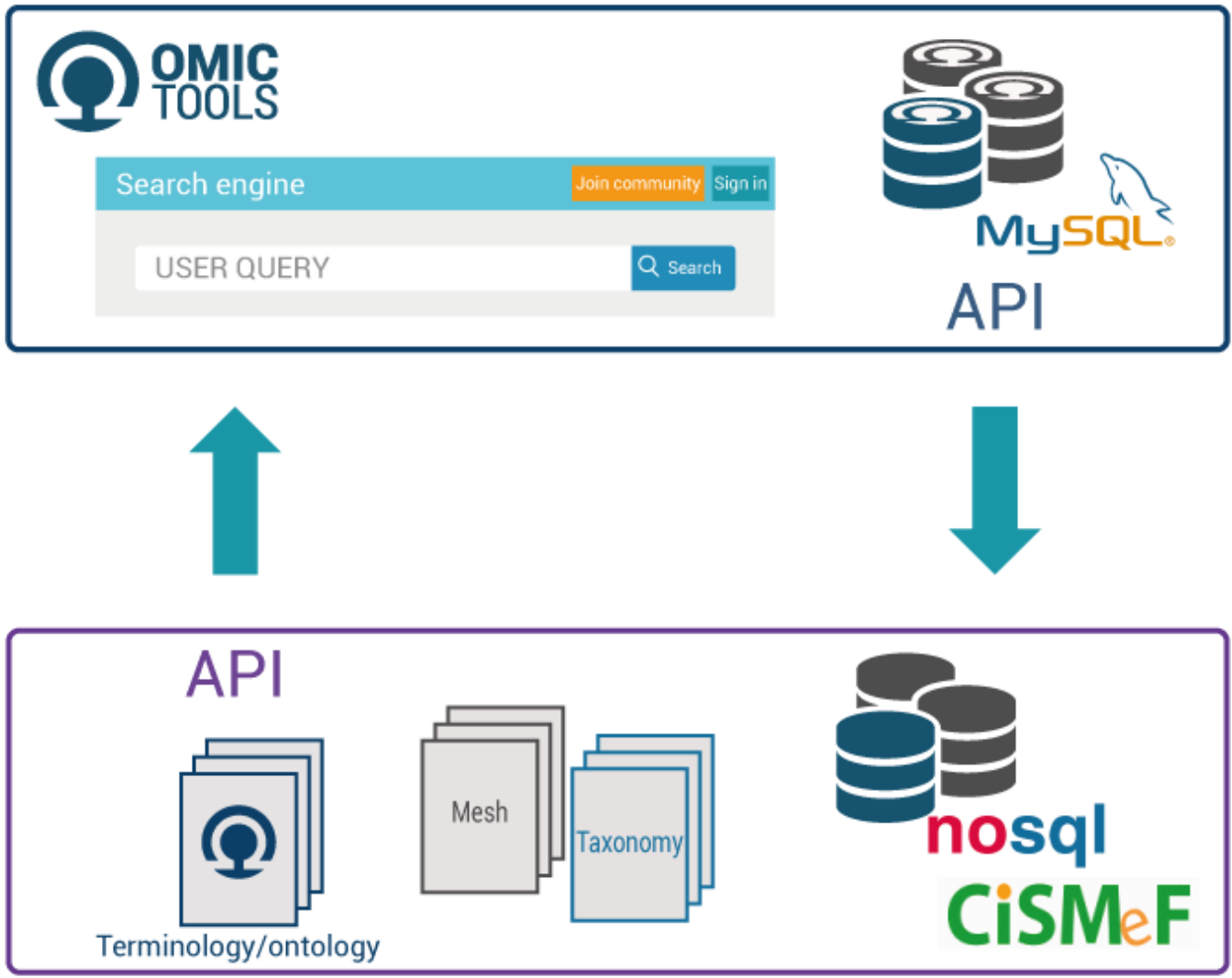

**Suppl. Figure 2**